\documentclass[trackchanges,twocolumn]{aastex701}

\usepackage{multirow}

\begin{document}

\title{Magnetohydrodynamical opening of dust traps in protoplanetary disks}

\author[orcid=0000-0002-4439-6831,sname='Khaibrakhmanov']{Sergey Khaibrakhmanov}
\affiliation{Saint-Petersburg State University, 7/9 Universitetskaya emb., St. Petersburg, 199034, Russia}
\email[show]{s.khaibrakhmanov@gmail.com}  

\author[orcid=0000-0002-4324-3809,sname='Akimkin']{Vitaly Akimkin} 
\affiliation{Institute of Astronomy, Russian Academy of Sciences, Pyatnitskaya str. 48, Moscow, 119017, Russia}
\email{akimkin@inasan.ru}

\begin{abstract}

Observed ring-like structures in protoplanetary disks are often interpreted as local pressure maxima, which induce efficient dust concentration. We revisit this paradigm, considering the effect of the large-scale magnetic field stresses on the gas rotation speed. Our simulations show that the magnetic field can be dynamically strong and cause $1-2$\% deviation from the Keplerian rotation at the periphery of a typical turbulent disk with dust grains of size $> 1\,\mu$m. This effect increases the inward drift speed of large grains characterized by Stokes number of $0.01-0.1$ by up to two times in our simulations. Importantly, such MHD deviation from the Keplerian rotation does not depend on the local gas pressure gradient and leads to drift towards the star only. The fast drift induced by this effect can cancel out the outward drift caused by the positive pressure gradient at the inner edge of a ring and open up the dust trap. For the disks with turbulence parameter $\alpha=10^{-3}$, this effect appears in the rings with a half-width of $10$~au and a density contrast up to $60$\% ($200$\% for $\alpha=10^{-2}$). Thus, the presence of a large-scale magnetic field in protoplanetary disks either completely prevents or imposes stricter conditions for dust accumulation and the onset of the streaming instability in the density rings in protoplanetary disks.

\end{abstract}


\section{Introduction} 

The observational data indicate that young stars are born surrounded by non-uniform rotating gas-dust accretion disks threaded by a large-scale magnetic field. 
The rotation of such disks is mainly Keplerian, but the deviations from the Keplerian speed of the order of $1$\% are measured~\citep{oberg2021, fukagawa2026}. The largest deviation is observed at the periphery of the disks. The magnetic field of the disks has complex geometry as revealed through polarization mapping of dust thermal emission in a few objects~\citep{li2016, li2018, thang2024, lietzow2025, ohashi2025}. Measurements of the magnetic field strength are still challenging~\citep[see review][]{kh24}, but the study of the Zeeman effect is a promising perspective~\citep{teague2025}.

The accretion disks of young stars are believed to evolve into protoplanetary disks, in which planets can form. This assumption is in particular supported by a variety of annular substructures, which are often interpreted as dust density enhancements and found in most gas-dust protoplanetary disks~\citep{huang2018}. The rings and gaps are detected across a wide range of radii, from within $5$~au to over $150$~au, with typical widths scaling between $0.05$ and $0.2$ of the local radius. Millimeter-wave continuum observations are $\sim 10$ times more sensitive to dust grains with $0.1-1$~mm sizes, which are settled to the midplane~\citep{pavyar2019}. 
There are indications that the rings are characterized by local deviations from Keplerian rotation~\citep{oberg2021, fukagawa2026}.

Several mechanisms have been proposed to generate the annular structures~\citep[see recent review][]{drazkowska2023}: gravitational interaction with embedded protoplanets, snowlines, viscosity transitions at the boundary of the regions with damped turbulence and reduced accretion speed (dead zones), and MHD-induced zonal flows. 
These ring-like structures are most commonly interpreted as dust traps inside local pressure maxima, since the vanishing radial pressure gradient inside the ring eliminates the gas drag that drives orbital drift of solid particles. This allows the particles to accumulate in the rings and reach dust-to-gas ratios approaching or exceeding unity.  Such high local concentrations of dust are essential for overcoming the `fragmentation barrier' and triggering the streaming instability, which leads to the direct gravitational collapse of pebble clumps into planetesimals. 

It remains unclear how the mechanism of dust traps may change in the presence of the magnetic field. Theoretical models have shown that the magnetic fields influence the structure of the disk in many ways: the magnetic field is important for MHD turbulence generation via the magnetorotational instability (MRI), it can also can drive strong outflows and corresponding angular momentum transport~\citep[see review by][]{lesur2023}; magnetic fields change the vertical structure of the disk~\citep{lovelace2009, lizano2016, khd22}, influence its temperature~\citep{lizano2016, khd2019mhd, mori2019, bethune2020} as well as the velocity field~\citep{shu2007, khd22}. In particular, magnetic tensions contribute to the centrifugal equilibrium in disks and cause deviations from the Keplerian rotation. The latter fact allows us to hypothesize that the magnetic field could change the picture of solid particles drift in protoplanetary disks.

In this paper, we use the model of~\cite{khd22} (Section~\ref{sec:model}) to study the impact of the magnetic field on dust trapping.  We calculate the structure of a typical protoplanetary disk (Appendix~\ref{app:disk_structure}) and analyze the conditions for the generation of a dynamically strong magnetic field (Section~\ref{sec:MF}), which can influence the rotation profile. Then, we apply the resulting structure of the disk with a superimposed ring-like density bump to calculate the speed of the radial drift of dust grains (Section~\ref{sec:subkepler}).

\section{Model of the accretion disk}
\label{sec:model}

To calculate the structure of the disk with a large-scale magnetic field, we use a model presented in~\cite{khd22}. It is based on the approximations of the standard model of~\cite{ss1973}. In addition to the standard equations of that basic model, we self-consistently solve the induction equation in a stationary approximation. The model allows for an accurate calculation of the magnetic field components, taking Ohmic dissipation, magnetic ambipolar diffusion, and magnetic buoyancy into account. The detailed description of the model is given in Appendix~\ref{app:model}.

The main dynamical effect which we consider is the contribution of the large-scale magnetic tensions $\propto B_rB_z$ to the centrifugal equilibrium described by Eq.~(\ref{eq:Omega}). If we additionally take the effect of radial gas pressure into account~\citep{weidenschilling1977}, then Equation~(\ref{eq:Omega}) can be rewritten as:
\begin{equation}
v_\varphi=\sqrt{\frac{GM}{r}-\frac{r}{\rho}\frac{\partial P}{\partial r}-\frac{rB_rB_z}{2\pi\Sigma}}.\label{eq:v_phi}
\end{equation}
It is convenient to represent Equation~(\ref{eq:v_phi}) in a non-dimensional form~\citep[see][]{khd22}:
\begin{equation}
v_\varphi=v_{\rm K}\sqrt{1-\beta_{\rm K}^{\rm HD}-\beta_{\rm K}^{\rm MHD}},
\label{eq:v_phi2}
\end{equation}
where
\begin{eqnarray}
\beta_{\rm K}^{\rm HD} &=& \Bigg(\frac{1}{\rho}\frac{\partial P}{\partial r}\Bigg)\frac{1}{g_r},\label{eq:beta_k_GD}\\
\beta_{\rm K}^{\rm MHD} &=& \Bigg(\frac{B_rB_z}{2\pi\Sigma}\Bigg)\frac{1}{g_r}
\label{eq:beta_k_MGD}
\end{eqnarray}
are the degrees of the deviation of the gas velocity from the Keplerian one due to the action of the gas pressure gradient and magnetic tensions, respectively. We used the definition of gravitational acceleration $g_r = GM/r^2$ in Equations~(\ref{eq:beta_k_GD}--\ref{eq:beta_k_MGD}). Expression~(\ref{eq:v_phi2}) shows that the difference between the Keplerian velocity and $v_\varphi$ is $\Delta v=v_{\rm K}-v_\varphi \approx1/2\beta_{\rm K}v_{\rm K}$, where $\beta_{\rm K}$ is one or both of the quantities from (\ref{eq:beta_k_GD}--\ref{eq:beta_k_MGD}). For example, one obtains $\Delta v\approx1\%v_{\rm K}$ in the case when $\beta_{\rm K}=0.02$.

\section{Results}
\label{sect:results}

\subsection{Model parameters}
\label{sec:params}

We model the disk structure using the standard parameters for a typical T~Tauri star: stellar mass $M_\star=1\,M_\odot$, stellar radius $R_\star=2\,R_\odot$, stellar luminosity $L_\star=2\, L_\odot$, magnetic field strength at the surface of the star $B_\star=2$~kG, accretion rate $\dot{M}=10^{-8}M_\odot\,\text{yr}^{-1}$, turbulence parameter $\alpha=0.01$.

We adopt the following standard parameters for ionization model: dust grain radius $a_{\rm d}=0.1\,\mu$m, unattenuated ionization rate by cosmic rays $\xi_{\rm CR0}=10^{ -17}$~s$^{-1}$, cosmic rays attenuation depth $\Sigma_{\rm CR} = 96$~g~cm$^{-2}$, X-ray luminosity of the star $L_ {\rm XR}=10^{30}$~erg~s$^{-1}$, ionization rate due to the decay of radionuclides $\xi_{\rm RE}=7.6\cdot10^{-19}$~s$^{-1}$.
Following~\cite{sano2000}, the characteristic value of the metals' abundance is assumed to be $X_{\rm a}=7.97\cdot 10^{-5}$.  

\subsection{Magnetic field strength}
\label{sec:MF}
Let us search for the conditions under which a dynamically strong magnetic field can be generated in the disk. We define the dynamically strong field in such a way that the corresponding plasma beta is equal to or smaller than unity. The plasma beta is defined in the midplane of the disk as: $\beta=8\pi\rho_0c_{\rm T}^2/B_z^2$.
The magnetic field strength is determined mainly by the efficiency of Ohmic dissipation and magnetic ambipolar diffusion~\citep{dkh2014}. In turn, the ionization fraction is the main parameter that determines the efficiency of diffusion processes.  

The radial distribution of the ionization fraction in the disk depends on corresponding density and temperature distributions (see Appendix~\ref{app:disk_structure}). Typically, the ionization fraction is large near the inner edge of the disk ($x\sim 10^{-10}$), very small ($x\sim 10^{-15}-10^{-12}$) in the intermediate region (the dead zone) and large at the disk periphery ($x\sim 10^{-12}-10^{-10}$).  The ionization fraction and extent of the dead zone depend strongly on the dust grain size. The dead zone has the size $0.2<r<8$~au in the case when grains have sizes $a_{\rm d}=0.1\,\mu$m and is absent in the case $a_{\rm d}=1$~mm.

\begin{figure}
  \centering
  \includegraphics[]{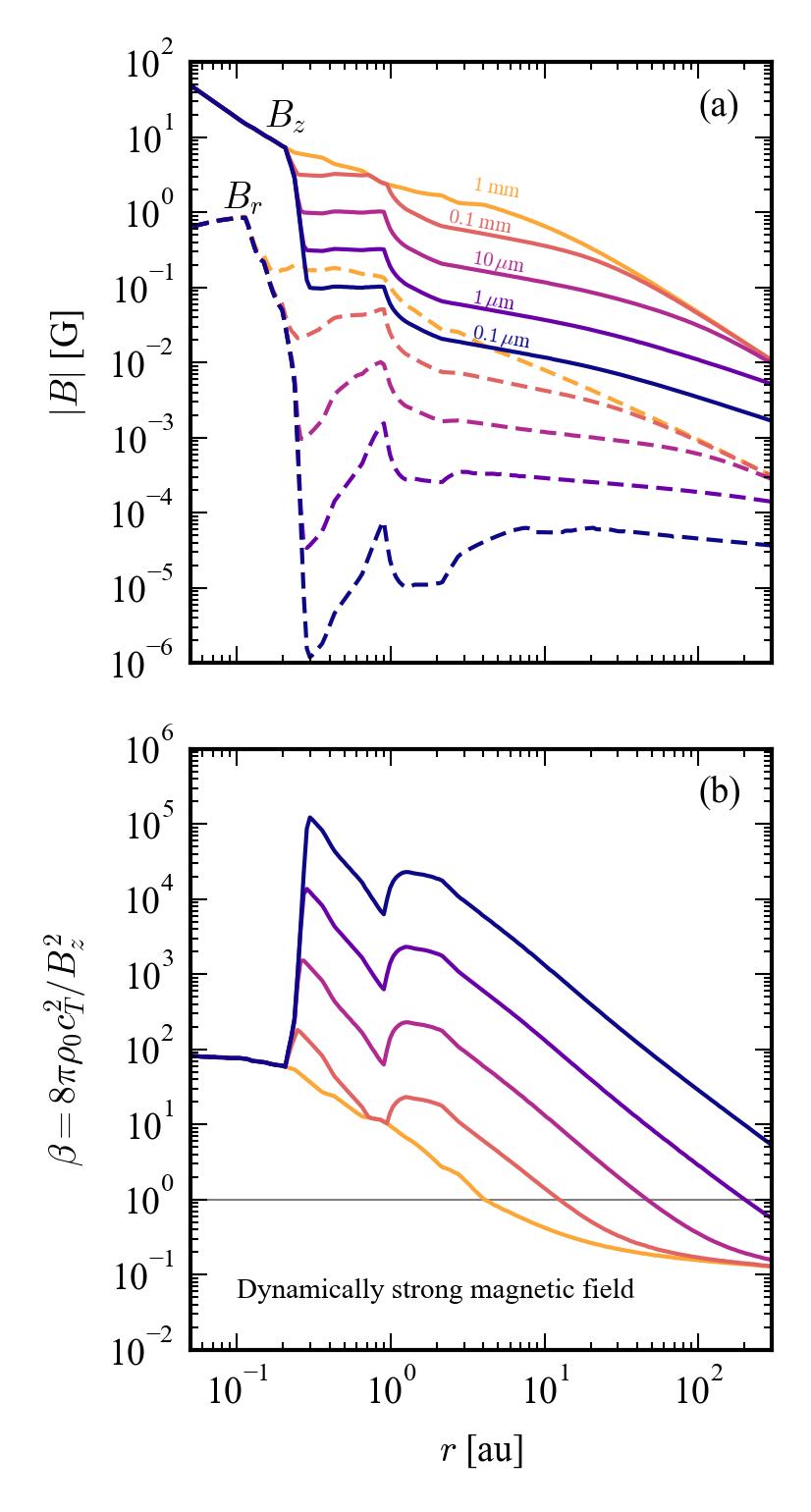}
  \caption{Panel (a): the radial profiles of the magnetic field components $B_r$ (dashed lines) and $B_z$ (solid lines) for various dust grain sizes (lines of different colors). Panel (b): corresponding radial profiles of plasma $\beta$; the horizontal line delineates value $\beta=1$.}
\label{fig:B_vs_ad}
\end{figure}

Such a non-uniform distribution of the ionization fraction shapes the radial profiles of the magnetic field components in the disk. In Figure~\ref{fig:B_vs_ad}, we plot the radial profiles of the magnetic field components $B_r$ and $B_z$, as well as the plasma $\beta$, for various dust grain sizes. The plasma beta is defined in the midplane of the disk as: $\beta=8\pi\rho_0c_{\rm T}^2/B_z^2$.  
We do not show the values of $B_\varphi$ here, since our aim is to study the effect of magnetic tensions $\propto B_rB_z$ in this paper. The detailed analysis of the relation between all three magnetic field components can be found in \cite{kh2017a}.

Figure~\ref{fig:B_vs_ad}a shows that, starting from the strength of $500$~G at the inner disk edge, the value of $B_z$ monotonically decreases and changes proportionally to gas surface density $\Sigma$ in the region of thermal ionization, $r<0.2$~au. The magnetic field is tightly coupled to plasma due to a large ionization fraction here. The value of $B_z$ abruptly falls by a factor of $2-100$ (depending on dust grain size) at $r\approx0.2-0.3$~au. This drop marks the transition to the dead zone, where the magnetic diffusion sets in. The region $r=(0.3-1)$~au inside the dead zone is characterized by a nearly constant value of $B_z$ ranging from $\approx 0.1$~G ($a_{\rm d} =0.1\,\mu$m) to $5$~G ($a_{\rm d}= 1$~mm). Outside this region of grain ice mantle evaporation, the profile $B_z(r)$ is a gradually decreasing function of the radial distance. The dependence of $B_z$ on $a_{\rm d}$ is explained by the fact that larger values of $a_{\rm d}$ correspond to larger values of the ionization fraction (see Figure~\ref{fig:disk_fiduc}c) and, correspondingly, to less efficient magnetic ambipolar diffusion, since the ambipolar diffusivity $\eta_{\rm MAD}\propto x_{\rm e}^{-1}$ and $x_{\rm e}\propto a_{\rm d}$. In the case $a_{\rm d}=1$~mm, the dead zone is absent, and the magnetic field is tightly coupled to the plasma throughout the entire disk, $B_z\propto \Sigma$.

In the absence of the dead zone ($a_{\rm d}=1$~mm), the radial profile of $B_r$ resembles that of $B_z$ with $B_r\sim 10^{-2}\,B_z$.
When $a_{\rm d}\leq 0.1$~mm, the value of $B_r$ inside the dead zone, $r<0.3$~au and $r\gtrsim 10$~au, is reduced by $4-5$ orders of magnitude, which is the result of the Ohmic dissipation. The value of $B_r$ increases with dust grain size, similar to $B_z$ inside the dead zone.

Figure~\ref{fig:B_vs_ad}b shows that the radial profile $\beta(r)$ appears as the inversed profile of the magnetic field components. It has the smallest value of $100$ in the innermost region of the disk. If $a_{\rm d}<1$~mm, the plasma beta rapidly grows with distance at the inner boundary of the dead zone, $r=(0.2-0.3)$~au, reaching a maximum value of $10^5$ ($a_{\rm d}=0.1\,\mu$m) to $200$ ($a_{\rm d}=0.1$~mm). The disk with the largest grains, $a_{\rm d}=1$~mm, is characterized by the smallest value of plasma beta, since the magnetic field strength has the maximum possible value in this case (see Figure~\ref{fig:B_vs_ad}a). The value of $\beta$ decreases with distance outside the dead zone because the gas pressure, $p=\rho c_{\rm T}^2$, decreases with distance faster than the magnetic pressure, $p_{\rm m}=B_z^2/8\pi$.  
The plasma beta falls below unity in the outer part of the disk if $a_{\rm d}> 1\,\mu$m. The inner radius of this region ranges from $\sim 4$~au ($a_{\rm d}=1$~mm) to $\sim50$~au ($a_{\rm d}=10\,\mu$m). Therefore, the periphery of the disks with large dust grains is characterized by the dynamically strong magnetic field, which can influence the structure of the disk. 

The presence of a density ring changes both the ionization and magnetic field structure. This effect is described in details in Appendix~\ref{app:x_B}. The increased density inside the ring leads to decrease in the ionization fraction, since the value of $x$ is inversely proportional to gas density (see Figure~\ref{fig:x_B_ring}(a)). The distribution of the magnetic field inside the ring depends on the dust grain size, as Figure~\ref{fig:x_B_ring}(b) demonstrates. In the case of large grains, $a_{\rm d}\geq 10\,\mu$m, the magnetic field is tightly coupled to gas and its strength increases inside the ring proportionally with surface density. In the case of smaller grains, efficient magnetic ambipolar diffusion decreases magnetic field strength inside the ring.

\subsection{Sub-Keplerian rotation}
\label{sec:subkepler}

\begin{figure*}
  \centering
  \includegraphics[]{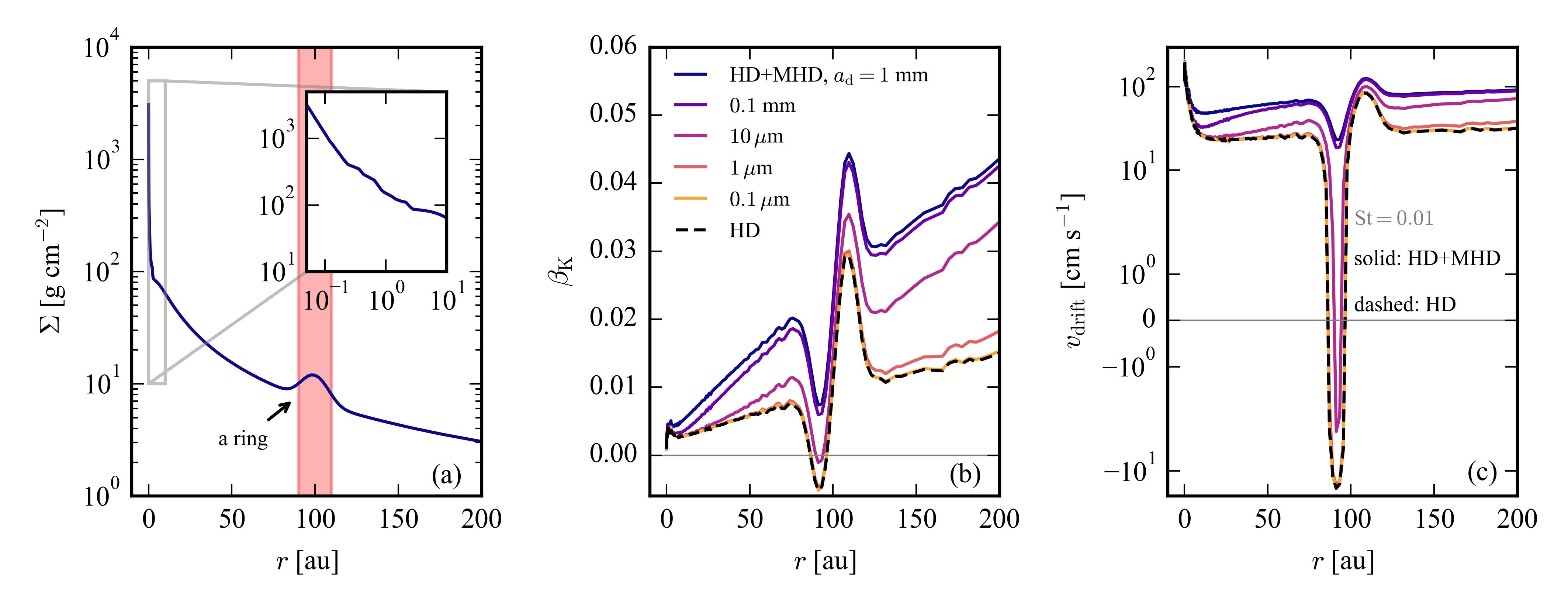}
  \caption{Panel (a): the radial profile of the surface density of the disk with a ring-like density bump at $R_{\rm ring}=100$~au. The ring has a half-width $10$~au and a maximum density $\Sigma_{\rm bump} = 2\times \Sigma_{\rm disk}(R_{\rm ring}) \approx 12$~g~cm$^{-2}$. The inset shows the zoomed-in inner region of the disk. Panel~(b): corresponding radial profiles of the degree of the deviation of plasma rotation speed from the Keplerian one, $\beta_{\rm k}$, calculated for various dust grain sizes. Dashed line: the HD case when the deviation is caused by a radial gas pressure gradient. Solid lines: the case when both HD and MHD deviations are considered. Panel~(c): corresponding radial profiles of the stationary speed of the radial drift for particles with Stokes number ${\rm St}=0.01$.}
\label{fig:subkepler_ring}
\end{figure*}

To study the relative role of the hydrodynamic (HD) deviation term~(\ref{eq:beta_k_GD}) and the MHD deviation term~(\ref{eq:beta_k_MGD}) in Equation~(\ref{eq:v_phi2}), we consider a disk with a ring-like density bump $\Sigma_{\rm ring}$ superimposed on the monotonic background $\Sigma_{\rm disk}$
\begin{equation}
\Sigma_{\rm ring}(r) = \Sigma_{\rm bump}\exp{\left[-\frac{\left(r - R_{\rm ring}\right)^2}{\Delta R_{\rm ring}^2}\right]},\label{eq:ring_density}
\end{equation}
where the density ring is supposed to have a radius $R_{\rm ring}$, half-width $\Delta R_{\rm ring}$ and maximum surface density $\Sigma_{\rm bump}$. The half-width of the ring corresponds to the standard Gaussian deviation times~$\sqrt{2}$. We take $R_{\rm ring}=100$~au, $\Delta R_{\rm ring}=10$~au and $\Sigma_{\rm bump}=2\times \Sigma_{\rm disk}(R_{\rm ring})$ as standard values. Here the ring width is taken as the maximum value according to observational constraints from~\cite{huang2018}: $\Delta R_{\rm ring}=0.1\,R_{\rm ring}$. The opposite case of a narrow ring is considered in Appendix~\ref{app:sharper_ring}. The total surface density of the disk is calculated as $\Sigma(r) = \Sigma_{\rm disk} + \Sigma_{\rm ring}$, where $\Sigma_{\rm disk}$ is a result of the simulation of the disk structure using the model presented in Section~\ref{sec:model} (depicted in Figure~\ref{fig:disk_fiduc}a).

In order to analyze how the HD and MHD deviations influence the dust grain radial drift, we calculate the corresponding radial drift speed in a stationary approximation for the case of Epstein drag~\citep[see, e.g.][]{armitage_book},
\begin{equation}
    v_{\rm drift} \approx -\beta_{\rm K}\mathrm{St}v_{\rm K},\label{eq:v_drift}
\end{equation}
where St is the Stokes number defined as a particle stopping time expressed in units of the local Keplerian timescale $\Omega_{\rm K}^{-1}$. The Equation~(\ref{eq:v_drift}) expresses the relative velocity of the particle with respect to the gas. Thus, the drift of small particles ($\mathrm{St}\ll1$) together with the gas (so-called advective drift) in the process of gas accretion is excluded from consideration since it does not depend on the peculiarities of the gas rotation law.

We consider two populations of dust particles in our simulations. The first one represents the dust which is the well-mixed with gas and acts as a source of opacity and electron recombinations affecting the ionization fraction. The drift of these dust grains is not considered. We vary the size of these dust grains, as in Figure~\ref{fig:B_vs_ad}, to study the changes in the magnetic field distribution related to the corresponding changes in the ionization fraction. The second one is the population of drifting particles moving through the disk with the speed defined by Equation~(\ref{eq:v_drift}). The size of such drifting particles is defined by the adopted value of the Stokes number.

In Figure~\ref{fig:subkepler_ring}, we plot the radial profiles of the gas surface density in the disk with a ring-like density bump, the corresponding degrees of HD and MHD deviations of the gas rotation speed from the sub-Keplerian one, $\beta_{\rm K}$, and the stationary speed of the radial drift of dust grains. We consider various sizes of non-drifting dust grains since the MHD deviation depends on the magnetic field strength, which is influenced by the value of $a_{\rm d}$. For clarity, we consider the drifting particles to be described by a fixed Stokes number $\rm{St}=0.01$. The case with the Stokes number determined by the dust fragmentation is considered in is considered in Appendix~\ref{app:stokes}. The simulations show that our general conclusion regarding the impact of the magnetic field on the drift speed remains the same in this case.

Figure~\ref{fig:subkepler_ring} shows that the HD deviation $\beta_{\rm K}^{\rm HD}$ linearly increases with distance in the inner part of the disk, ranging from $\sim 10^{-3}$ at the inner radius of the disk to $\approx 0.005$ at $r\sim 80$~au. The periphery of the disk is also characterized by a linear radial profile of the deviation $\beta_{\rm K}^{\rm HD}$, ranging from $0.01$ at $r=130$~au to $\sim 0.015$ at the outer radius of the disk. The dependence of $\beta_{\rm K}^{\rm HD}$ on $r$ is non-monotonic inside the ring. This dependence is antisymmetric with respect to the center of the ring: $\beta_{\rm K}^{\rm HD}$ has a minimal value of $(-5\cdot 10^{-3})$ at $r\approx 90$~au (at the inner ring edge) and a maximum value of $0.03$ at $r\sim 110$~au (at the `right' ring edge). The value of $\beta_{\rm k}^{\rm HD}$ changes sign across the ring, which is a direct consequence of the changing sign of the radial gradient of the gas pressure.

Comparison of the $\beta_{\rm K}(r)$ profiles calculated taking both HD and MHD deviations shows that the contribution of the magnetic tensions increases with the size of non-drifting particles $a_{\rm d}$. When $a_{\rm d}=0.1\,\mu$m, the effect of the magnetic tensions is negligible, $\beta_{\rm K}\approx \beta_{\rm K}^{\rm HD}$, because the magnetic field is dynamically unimportant under such conditions (see Figure~\ref{fig:B_vs_ad}). The MHD deviation manifests itself in the cases of $a_{\rm d}\geq 1\,\mu$m, when the magnetic field becomes dynamically strong (see Figure~\ref{fig:B_vs_ad}b). The effect of the magnetic tensions leads to large values of $\beta_{\rm K}$ at the periphery of the disk, including the region of the ring.  In the case of $a_{\rm d}=1$~mm, the total deviation from the Keplerian rotation is of $0.02$ at $r=75$~au ($\Delta v\sim 1$\%~$v_{\rm K}$, which is $\sim 2.5$ times larger than the HD deviation) and of $0.045$ at $r=110$~au ($\Delta v\sim 2.25 $\%~$v_{\rm K}$, i.~e. $1.5$ times larger than $\beta_{\rm K}^{\rm HD}$). The maximum value of the deviation is reached at the outer edge of the disk, $\beta_{\rm K}\approx 0.043$ ($\Delta v\approx 2.15$\%~$v_{\rm K}$). The most important result is that the value $\beta_{\rm K}$ does not change sign across the ring when the magnetic stresses are taken into account. Thus, the magnetic field effectively `dismantles' the dust trap.

The differences between the HD and MHD deviations have a drastic effect on the speed of the radial drift of dust grains, as Figure~\ref{fig:subkepler_ring} shows. The case of purely HD deviation demonstrates a classical picture of a dust trap inside the ring. The radial profile of $v_{\rm drift}$ is characterized by nearly constant values of $25$~cm~s$^{-1}$ in the inner region of the disk, $r\lesssim 75$~au, and $\sim 30$~cm~s$^{-1}$ at the disk's periphery, $r\gtrsim 120$~au. The profile of $v_{\rm drift}(r)$ is non-monotonic inside the ring.  The change of the sign of $\beta_{\rm K}^{\rm HD}$ across the ring switches the direction of the radial drift (i.~e. $v_{\rm drift}$ also changes sign; see Equation~\ref{eq:v_drift}). This causes the dust grains to concentrate inside the ring.

Figure~\ref{fig:subkepler_ring}c shows that the picture qualitatively changes when the MHD deviation from the Keplerian rotation becomes prominent. First of all, radial drift velocity becomes proportionally larger with $\beta_{\rm K}$. In the case of $a_{\rm d}=1$~mm, the drift speed is nearly two times greater than in the HD case: $v_{\rm drift}\approx 50-70 $~cm~s$^{-1}$ in the inner region of the disk and $\approx 80-95$~cm~s$^{-1}$ at the disk's periphery. Second, the radial drift speed does not change sign across the ring when $a_{\rm d}\geq 1\,\mu$m. The radial profile of $v_{\rm drift}$ is characterized by a decreased value inside the ring, but this value does not become negative.

\section{Discussion}
\label{sect:discuss}

We found that the poloidal magnetic stresses can cause the gas rotation speed to deviate from the Keplerian one by $1-2$~\% at the periphery of the disk. This MHD deviation dominates over the HD one, caused by the gas pressure gradient, in disks with slightly grown dust grains, $a_{\rm d}\geq 1\,\mu$m, when the magnetic field becomes sufficiently large due to reduced magnetic diffusivity.  
Our results agree well with recent ALMA high-spectral observations of several protoplanetary disks~\citep{oberg2021, teague2021, teague2022, fukagawa2026}. Usually, the deviation from the Keplerian speed is attributed to the gas pressure gradient~\citep[see][]{pinte2023}. Our simulations show that the MHD deviation can be a dominant mechanism at the periphery of the disks with evolved dust grains. The MHD deviation depends on the local ionization state and is unlikely to operate inside the dead zones of reduced magnetic field strength. We propose that the study of an `inhomogeneity' of the sub-Keplerian rotation inside the disks can be an indirect tool to analyze the magnetic fields in disks. 

A remarkable feature of the MHD deviation from the Keplerian rotation, $\beta_{\rm K}^{\rm MHD}$, is that it always leads to a gas deceleration (considering the classical picture of the magnetic field distribution with poloidal magnetic field lines bent away from the star). In contrast, the effect of the radial gas pressure gradient, $\beta_{\rm K}^{\rm HD}$, can lead to both deceleration of the gas rotation (pressure increases towards the star) and its acceleration (pressure decreases towards the star). The competition between these two effects can lead to the following scenarios. If $\beta_{\rm K}^{\rm HD}>0$ then the MHD deviation increases the speed of the drift towards the star. If $\beta_{\rm K}^{\rm HD}<0$, then the direction of the drift can be changed by the effect of the magnetic tensions given that $\beta_{\rm K}^{\rm MHD} > |\beta_{\rm K}^{\rm HD}|$. The latter scenario can be realized inside a local pressure maximum in the disk.
Our simulations demonstrate that, for the disk of a typical T~Tauri star with a density ring placed at $r=100$~au and having a width of $20$~au and peak density of $12$~g~cm$^{-2}$ (two times larger than the surrounding disk density), the MHD deviation dominates over the HD one if non-drifting dust grains have a radius larger than $1\,\mu$m. Under such conditions, the total deviation from the Keplerian rotation does not change sign across the ring, and the pressure maximum does not act as a dust trap. 

The MHD deviation from the Keplerian rotation depends on the magnetic field strength, which, in turn, depends on the rate of magnetic field advection in the disk. This effect is controlled by the turbulence parameter in our simulations. Smaller values of $\alpha$ cause a smaller magnetic field strength. This effect is studied in Appendix~\ref{app:alpha_dependence}. Our simulations show that the disks with moderate turbulence ($\alpha =10^{-3}$) are still characterized by dynamically important magnetic field at the disk's periphery. In this case, the MHD deviation from the Keplerian rotation can become sufficiently large to open the dust trap in the case of wide rings with low density contrast only: $\Delta R_{\rm ring}=10$~au and $\Sigma_{\rm ring}/\Sigma_{\rm disk}(R_{\rm ring})=0.6$. The disks with low turbulence levels ($\alpha=10^{-4}$) have a kinematic magnetic field (plasma $\beta>1$), which cannot open the dust trap for the considered values of $\Delta R_{\rm ring}=(0.25-10)$~au and $\Sigma_{\rm ring}/\Sigma_{\rm disk}(R_{\rm ring})=(0.1-10)$.

It should be noted that the effect of `dust trap opening' by the magnetic tensions depends on the parameters of the ring. For example, our simulations performed for the highly turbulent disk with a sharper pressure maximum with $\Delta R_{\rm ring}=0.05\,R_{\rm ring}$ (see Appendix~\ref{app:sharper_ring}) showed that the effect of the magnetic tensions is insufficient to open the dust trap in such cases, and the local pressure maximum still halts the radial drift. However, a shallower pressure maximum (smaller $\Sigma_{\rm ring}$) favors the MHD opening of a trap. A parameter study presented in Appendix~\ref{app:ring_dependence} shows that the wide rings with $\Delta R_{\rm ring}=0.1\,R_{\rm ring}$ (maximum observed value) can be opened by the magnetic field in the case of a density contrast smaller than $200$~\%. The condition for the MHD opening of the sharpest rings ($\Delta R_{\rm ring}=0.05\,R_{\rm ring}$) is stricter: $\Sigma_{\rm ring}\leq 20\,\%\,\Sigma_{\rm disk}(R_{\rm ring})$. 

In our study, we neglected the effect of the vertical magnetic stresses and MHD outflows on the angular momentum transport in the disks. This effect has now been widely studied under a wide range of conditions~\citep[see review by][]{lesur2023}. We emphasize that self-consistent modeling of the disks with MHD outflows requires global modeling that connects the disk with its atmosphere, which goes beyond the capabilities of our MHD model of the disk. 
There are two main features caused by MHD winds in comparison to turbulence.
First, the MHD winds have been shown to decrease the size of the accretion disks~\citep{tabone2022}. Second, strong MHD winds can increase the gas accretion speed. This means that the \textit{advection} drift of dust particles, which are tightly coupled to gas ($\mathrm{St}\ll1$), will be faster in the disks with MHD winds. This effect can also change the drift speed through the local pressure maxima~\citep{wu2023, wu2024}. In our work, we considered the \textit{relative} drift of dust particles with respect to gas, which depends on the deviation of gas rotation speed from the Keplerian one and does not depend on the gas accretion speed. Therefore, the particle drift considered in our work does not depend on the specific mechanism of angular momentum transport. We propose that the effect of MHD opening of dust traps can operate both in turbulent disks and in laminar disks, in which the accretion is driven by MHD winds. This is an interesting topic for future studies. 

It should also be noted that sophisticated 3D MHD simulations of protoplanetary disks indicate that disks with magnetic fields often exhibit complex self-organization, actively forming zonal flows, localized flux accumulations, and time-dependent substructures~\citep{aoyama2023, WF2023, WF2025, hu2025}. Our (1+1)D MHD simulations allow for the consideration of radial magnetic field inhomogeneities only. The rings naturally introduce such inhomogeneities, as discussed in Appendix~\ref{app:x_B}. The particle drift in the presence of zonal flows and transient structures should be studied within a 3D approach.

\section{Conclusions}
Below, we summarize our main results and draw conclusions.

\begin{itemize}
 
    \item The regions of a dynamically strong magnetic field in protoplanetary disks are characterized by sub-Keplerian gas rotation. The deviation from the Keplerian law is due to large-scale magnetic stresses. This MHD deviation exceeds the effect of the radial gas pressure gradient in the disks with dust grains of size $a_{\rm d}\geq 1\,\mu$m and moderate to high turbulence $\alpha\geq 10^{-3}$ The difference between the gas rotation speed and the Keplerian speed reaches $2$\% at the periphery of the disk, $r>10-100$~au. We hypothesize that the sub-Keplerian rotation observed at the periphery of several protoplanetary disks~\citep{teague2021, teague2022} may be a sign of such an MHD effect. 
    \item The sub-Keplerian rotation in the regions of strong magnetic fields will cause faster radial drift of dust grains towards the star. 
    \item The deceleration of gas rotation caused by the magnetic tensions can cancel out the effect of large dust particles trapping inside local pressure maxima. This `dust trap opening' effect occurs when the magnetic tensions exceed the local radial gradient of gas pressure inside the pressure bumps (density rings) in the disk. The conditions for this effect of MHD opening of dust traps are favorable in the disk with $\alpha\geq 10^{-3}$, rings half-width~$>10$~au and density contrast up to~$60$\%. This novel effect poses a challenge for dust grain evolution in protoplanetary disks. The large-scale magnetic fields strongly limit the dust trapping inside the gas pressure maxima, degrading them to a 'traffic jam' ring. This imposes stricter conditions for the onset of the streaming instability.
    
\end{itemize}

\begin{acknowledgments}
We thank the anonymous referee for the deep and constructive comments, which helped present our results in more detail and clarify the limits of our approach. The work of Sergey Khaibrakhmanov is supported by the Theoretical Physics and Mathematics Advancement Foundation `BASIS' (project 23-1-3-57-1).
\end{acknowledgments}

\begin{contribution}

SK came up with the initial research concept, developed a numerical code based on the model presented, and was responsible for writing and submitting the manuscript. VA provided ideas for comparing the contributions of the HD and MHD effects inside the density rings. He also edited the manuscript.

\end{contribution}

\software{matplotlib \citep{matplotlib}.}

\appendix

\section{Model description}
\label{app:model}

To study the structure and dynamics of an accretion disk with a large-scale magnetic field, we use the system of MHD equations~\citep{LL8} taking the magnetic ambipolar diffusion into account~\citep[see][]{dkh2014}.

We consider a stationary, geometrically thin, and optically thick accretion disk consisting of gas-dust plasma. The mass of the disk is small compared to the mass of the star. We use cylindrical coordinates $(r,\,\varphi,\,z)$ and assume that angular momentum is transported in the radial direction $r$ through turbulence parametrized using a non-dimensional turbulence parameter $\alpha$ following~\cite{ss1973}. The disk is in centrifugal equilibrium along the $r$-coordinate and magnetostatic equilibrium in the vertical direction~$z$. The energy released in the disk due to viscous dissipation is carried away by radiation in the vertical direction. The velocity vector has components $\textbf{v}=\{v_r,\,v_\phi,\,v_z\}$, where $v_r$ is the accretion speed, $v_\varphi=\Omega r$ is the azimuthal plasma velocity, and $\Omega$ is the angular frequency. The magnetic field is described by a vector $\textbf{B}=\{B_r,\, B_\phi,\, B_z\}$.
In the approximation of the geometrically thin disk, we neglect radial derivatives with respect to the vertical derivatives in the model equations. 

Given the above approximations, the basic MHD equations can be split into the subsystems describing the radial and vertical disk structures. 
The radial structure of the accretion disk with a large-scale magnetic field is described by the following system~\citep[see details in][]{khd22}:

\begin{eqnarray}
\dot{M} & = & -2\pi rv_r\Sigma, \label{eq:M_dot}\\
v_\varphi & = & \sqrt{\frac{GM}{r}\Bigg(1+\frac{z^2}{r^2}\Bigg)^{-3/2}-\frac{rB_rB_z}{{\bf 2\pi}\Sigma}}, \label{eq:Omega}\\
\dot{M}\Omega f & = & 2\pi\alpha\Sigma c_{\rm T}^2, \label{eq:Transfer}\\
H & = & \frac{c_{\rm T}}{\Omega}, \label{eq:H}\\
\sigma_{\rm SB}T_{\rm eff}^4 & = &\frac{3}{8\pi}\dot{M}\Omega^2f + 2H\Gamma_{\rm CR}, \label{eq:T_eff}\\
T^4 &=& T_{\rm eff}^4\left(\frac{1}{2} + \frac{3}{8}\kappa_{\rm R}\Sigma\right),\label{eq:Tmid}\\
-zv_rB_z & = & B_r\eta, \label{eq:B_r}\\
-\frac{1}{2}\Omega_{\rm k}z^2\Bigg(3\frac{z}{r}B_z+B_r\Bigg) & = & B_\varphi\eta, \label{eq:B_phi}\\
B_z &=& \left\{ 
\begin{array}{rcl}
B_{z0}\frac{\Sigma}{\Sigma_0},\quad {R}_{\rm m} \gg 1,\\
\sqrt{4\pi x\rho^2 r|v_r|}, \quad {R}_{\rm m} < 1.\\
\end{array}
\right.\label{eq:Bz}
\end{eqnarray}
Equation~(\ref{eq:M_dot}) follows from the continuity equation, in which $\dot{M}$ is the accretion rate,
\begin{equation}
\Sigma=\int_{-H}^H\rho dz\approx2\bar{\rho} H
\end{equation}
is the gas surface density, $\bar{\rho}$ is the average density in the disk, $H$ is the accretion disk scale height defined in Eq.~(\ref{eq:H}) in accordance with the solution of the hydrostatic equilibrium equation for the disk characterized by the sound speed $c_{\rm T}$.

Equation~(\ref{eq:Omega}) is the solution to the centrifugal equilibrium equation, which takes the contribution of magnetic tensions into account. This equation shows that, in the absence of a magnetic field, the gas rotates at Keplerian speed:
\begin{equation}
    v_{\rm K} = \sqrt{\frac{GM}{r}} = \Omega_{\rm K}r.
\end{equation}
The second term from the right in Equation~(\ref{eq:Omega}) corresponds to magnetic tensions, which can, generally speaking, lead to both deceleration and acceleration of the gas, depending on the signs of the components $B_r$ and $B_z$.

The angular momentum transport equation (\ref{eq:Transfer}) is derived assuming that the main component of the viscous stress tensor is $\sigma^\prime_{r\varphi}$. Following~\citet{ss1973}, we estimate this component as $\sigma^\prime_{r\varphi}=-\alpha P$, where $\alpha$ is the dimensionless turbulent parameter. The value of $\alpha$ is defined as the relation between local sound speed and the typical speed of turbulent pulsations. The factor $f$ in Equation~(\ref{eq:Transfer}) comes from the inner boundary condition,
$$
f=1-(r_0/r)^{1/2},
$$
where $r_0$ is the radius of the inner boundary of the accretion disk. The factor $f$ accounts for zero viscous torque at the inner radius of the disk.

Equation~(\ref{eq:T_eff}) describes the balance between heating and cooling rates per unit disk surface with effective temperature~$T_{\rm eff}$. This equation takes viscous heating (first term on the right-hand side) and cosmic ray heating $\Gamma_{\rm CR}$ (second term on the right-hand side) into account. The latter is calculated following~\citet[][]{dalessio1998}. The influence of heating due to dissipative MHD effects was studied by~\citet{khd2019mhd} and is not considered in this paper for clarity. It is assumed that the energy released in the disk is carried away by radiation in the vertical direction. The temperature in the equatorial plane of the disk $T$ is determined from the solution of the radiative transfer Eq.~(\ref{eq:Tmid}) written in the diffusion approximation. The Rosseland mean opacity $\kappa_{\rm R}$ in Eq.~(\ref{eq:Tmid}) is calculated by interpolating the tables of~\citet{semenov2003} for low temperatures and OPAL model for high temperatures~\citep{OPAL}. The transition temperature between these two opacity regimes is $T=10^{3.75}$~K~$\approx 5600$~K.
Heating of the gas by stellar radiation also plays an important role in the outer parts of the disk. In this work, this effect is taken into account by adding additional terms to the solution of Eq.~(\ref{eq:Tmid}) according to the analytical solution from the work of~\citet{malbet1991}.

Equations~(\ref{eq:B_r})--(\ref{eq:Bz}) represent the radial, azimuthal, and vertical components of the induction equation. Equations~(\ref{eq:B_r}) and (\ref{eq:B_phi}) determine the magnetic field components $B_r$ and $B_\varphi$ at the height $z$ above the disk midplane. The values of $B_r$ and $B_\varphi$ are determined from the balance between the generation of $B_r$ and $B_\varphi$ due to differential rotation in corresponding directions, on one hand, and vertical diffusion of the magnetic field, on the other hand. The following natural symmetrical boundary conditions are assumed for these magnetic field components: $B_r(z=0)=0$ and $B_\varphi(z=0)=0$.

Equation~(\ref{eq:Bz}) is obtained for two special cases: a frozen-in magnetic field (magnetic Reynolds number~$R_{\rm m} = v_rH/\eta\gg 1$) and effective magnetic ambipolar diffusion ($R_{\rm m} < 1$). In the first case, $B_z\propto \Sigma$, and fossil magnetic field strength is determined by the boundary conditions $B_{z0}$ and $\Sigma_0$ at the outer boundary of the disk. The conditions are set in such a way that the magnetic field at the outer boundary coincides with the magnetic field of the surrounding protostellar cloud~\cite[see][]{dkh2014}. In the second case, $B_z$ is determined by the equality of the advection speed $v_r$ and the speed of the magnetic ambipolar diffusion.

Total magnetic field diffusivity is introduced in Eqs.~(\ref{eq:B_r})--(\ref{eq:B_phi}) as
\begin{equation}
\eta=\eta_{\rm OD}+\eta_{\rm AD} + v_{\rm B}H,
\end{equation}
where $\eta_{\rm AD}$ is the ambipolar diffusivity and $v_{\rm B}$ is the magnetic buoyancy speed~\citep[see][]{kh2017b}.

Magnetic diffusivities depend on the ionization fraction, which we calculate by solving the ionization-recombination balance equation~\citep[see][]{spitzer_BOOK} considering ionization by cosmic rays, stellar X-rays, and radioactive elements:
\begin{equation}
    (1-x)\xi = C_{\rm rr}x^2n_{\rm n} + C_{\rm eg}xn_{\rm n},\label{eq:x_DS}
\end{equation}
where $x=n_{\rm e}/n_{\rm n}$ is the ionization fraction; $\xi$ is the ionization rate; $C_{\rm rr}$ is the rate of radiative recombinations; $C_{\rm eg}$ is the rate of electron recombinations onto dust grains. 

The ionization rate by primary cosmic rays is calculated by taking cosmic ray fluxes from both sides of the disk into account~\citep[see, e.g.,][]{sano2000}. We also take the effect of the secondary electrons into account using an approximation of the numerical results of~\cite{2021ApJ...909..107I}. The ionization rate by stellar X-rays is calculated using the prescription by~\cite{glassgold1997}. The ionization rate caused by the decay of radionuclides is calculated in accordance with the work of~\cite{umebayashi2009}.

Equation~(\ref{eq:x_DS}) is formulated assuming that the contribution of dust grain charges to quasi-neutrality is negligible. This quadratic equation has a simple analytical solution, which takes the power-law form $x\propto n_{\rm n}^q$ in the limiting cases of radiative recombinations ($q=1/2$) and dust grain recombinations ($q=1$). \cite{dudsaz1987} introduced a simple yet effective way to take the dust grain evaporation into account. The geometrical cross-section of a dust grain appearing in the coefficient $C_{\rm eg}$ is calculated assuming that the dust grain radius depends on temperature. This dependence reflects the evaporation of dust layers with increasing temperature~\citep[see][]{dkh2014}.

We take the thermal ionization into account in the regions where gas temperature increases above $\approx 500$~K. We follow~\cite{dkh2014} and calculate the ionization fraction from Saha's equation, considering the thermal ionization of metals with low ionization potential and hydrogen. For $T\lesssim 5000$~K, it is sufficient to consider the ionization of potassium only, since this metal has the lowest ionization potential. In this case, the thermal ionization fraction
\begin{equation}
    x^{(\rm T)} = 1.8\cdot 10^{-11}\left(\frac{T}{1000\,\mathrm{K}}\right)^{3/4}\left(\frac{X_{\rm K}}{10^{-7}}\right)^{1/2}\left(\frac{n_{\rm n}}{10^{13}\,\mathrm{cm}^{-3}}\right)^{-1/2}\frac{\exp(-25000/T)}{1.15\cdot 10^{-11}},
\end{equation}
where $X_{\rm K}$ is the relative abundance of potassium.

\section{Structure of the accretion disk}
\label{app:disk_structure}
In this section, we describe the general structure of the disk, which is used to analyze its magnetic field in Section~\ref{sec:MF}. 

\begin{figure*}[t]
  \centering
  \includegraphics[]{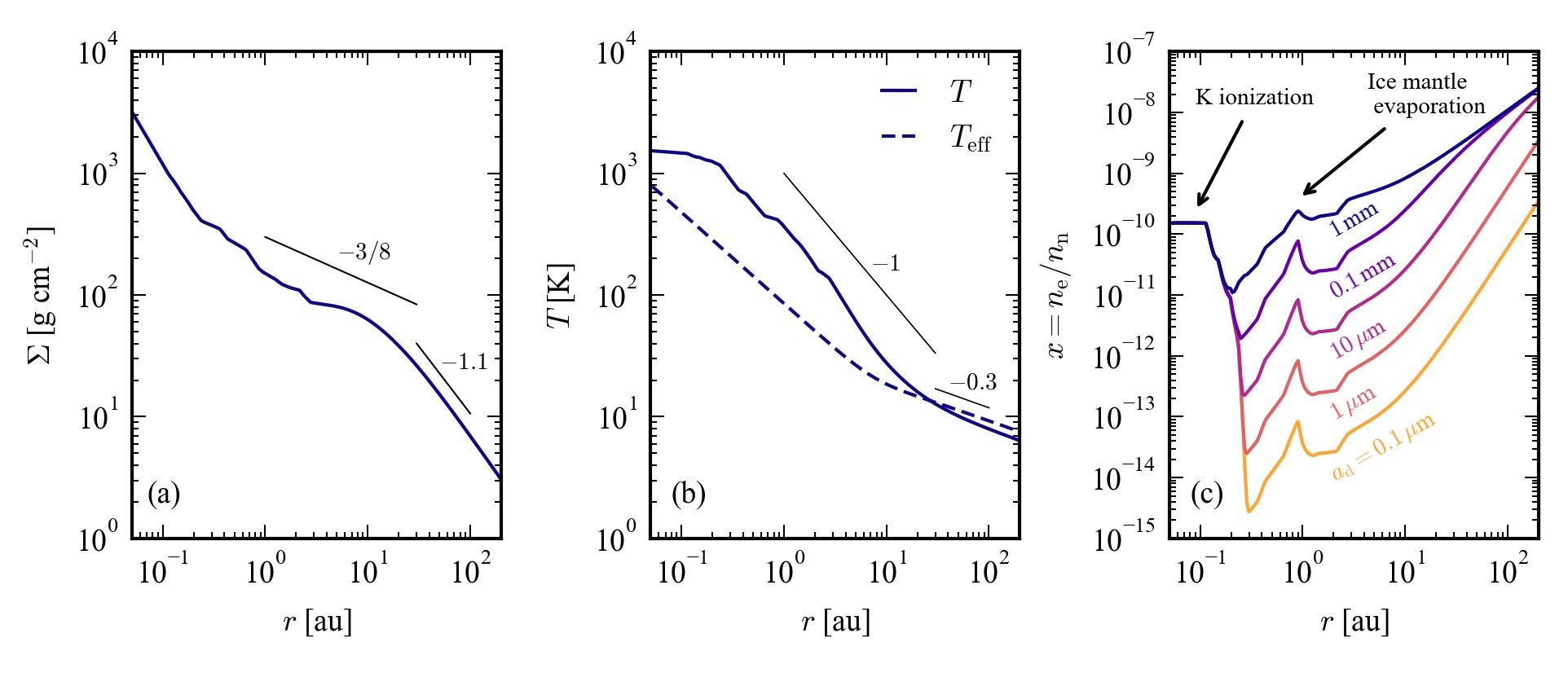}
  \caption{Radial structure of the disk. Panel~(a): surface density profile; panel~(b): the profiles of midplane temperature (solid line) and effective temperature (dashed line); panel~(c): midplane ionization fraction calculated for various dust grain sizes. Thin lines with labels show the slopes of the power-law profiles in corresponding regions.}
\label{fig:disk_fiduc}
\end{figure*}

In Figure~\ref{fig:disk_fiduc}, we plot the radial profiles of disk surface density and temperature, as well as the ionization fraction in the midplane of the disk.
Figure~\ref{fig:disk_fiduc}(a) and (b) demonstrate typical density and temperature structure of the disk with both $\Sigma$ and $T$ being nearly power-law functions of radial distance $r$. In the region $r>30$~au, the disk becomes optically thin relative to its own radiation, $T_{\rm eff}\approx T$. Changes in the slopes of the $\Sigma(r)$ and $T(r)$ profiles with distance are due to the changes in the opacity~\citep[see the analytical solution in][]{dkh2014}.

Figure~\ref{fig:disk_fiduc}(c) shows that the abundance of free electrons, $x_{\rm e}=n_{\rm e}/n_{\rm n}$, which approximately equals the ionization fraction of the plasma, is distributed along the radial distance in a very non-uniform way. It reaches a maximum value of $\sim 2\cdot 10^{-10}$ both at the inner and outer boundaries of the disk for $a_{\rm d}=0.1\,\mu$m case. This value is determined by thermal ionization of alkali metals in the inner region of the disk and cosmic ray ionization at its periphery. The intermediate region, $0.3<r<20$~au, has too low gas temperature for efficient thermal ionization and a sufficiently high gas surface density (cosmic rays are partly attenuated), and therefore the ionization fraction tends to be minimum, $x_{\rm min}\sim 10^{-15}$. This region corresponds to a dead zone, where magnetic diffusion quenches the MRI~\citep[][]{gammie1996}. A small peak in the $x_{\rm e}(r)$ profile at $r\approx 1$~au is due to evaporation of icy mantles of dust grains and the corresponding reduction of the rate of electron recombinations onto dust grains.

The ionization fraction varies proportionally to the dust grain size for $a_{\rm d}$ in the range from $0.1\,\mu$m to $1$~mm. This is due to the fact that the main recombination process under such conditions is the electron attachment to dust grains leading to $x_{\rm e}\propto a_{\rm d}$~\citep[see][]{dkh2014}. As a consequence, the extent of the dead zone, which can approximately be defined as a region of $x\lesssim 10^{-12}$, decreases. The dead zone in the disk with $a_{\rm d}\geq0.1$~mm is not formed. The boundaries of the dead zone calculated for various dust grain sizes are given in Table~\ref{tab:dead_zone}.

\section{Ionization fraction and the magnetic field in the disk with a ring}
\label{app:x_B}

\begin{figure*}[h!]
  \centering
  \includegraphics[]{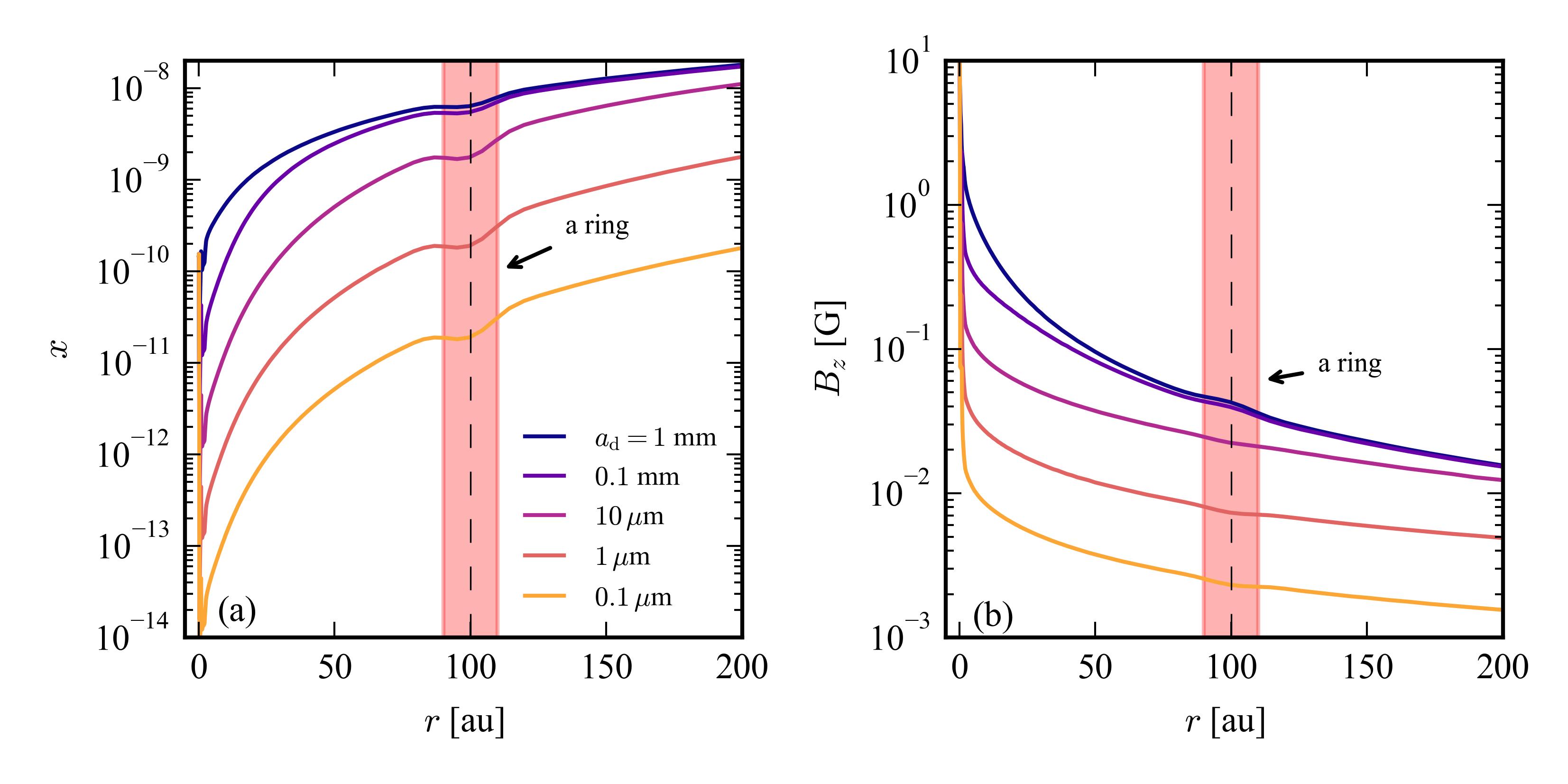}
  \caption{Radial profiles of the ionization fraction (a) and magnetic field strength (b) in the disk with a ring. Different lines correspond to various dust grain sizes. Vertical colored stripe delineates the ring position, the vertical dashed line shows the coordinate of the center of the ring. Standard disk and ring parameters are used. }
\label{fig:x_B_ring}
\end{figure*}

The MHD model of the disk, presented in Section~\ref{app:model}, allows for a self-consistent simulation of the structure of the disk with a ring. Therefore, the density ring introduced by Equation~(\ref{eq:ring_density}) not only changes the disk structure (i.e., density, temperature, velocity field), but also affects its ionization structure and the distribution of the magnetic field in the disk. To demonstrate this influence, we plot the radial profiles of the ionization fraction and the vertical magnetic field component in Figure~\ref{fig:x_B_ring}. We consider the disk with standard parameters and the ring described by a radius $R_{\rm ring}=100$~au, half-width $\Delta R_{\rm ring}=10$~au and maximum surface density $\Sigma_{\rm bump}=2\times \Sigma_{\rm disk}(R_{\rm ring})\approx 12 \,\mathrm{g}\,\mathrm{cm}^{-2}$. Both the ionization fraction and the magnetic field are calculated for various dust grain sizes.

\begin{table}[h!]
    \centering
    \begin{tabular}{|c|c|c|}
        \hline
        $a_d [\mu$m] & $r_{\text{in}}^{\text{(DZ)}}$ [au] & $r_{\text{out}}^{\text{(DZ)}}$ [au] \\
        \hline
        $0.1$ & $0.22$ & $7.88$ \\
        $1$ & $0.22$ & $2.49$ \\
        $10$ & $0.22$ & $0.50$ \\
        $100$ & $0.24$ & $0.26$ \\
        $1000$ & - & - \\
        \hline
    \end{tabular}
    \caption{Dependence of the dead zone inner ($r_{\rm in}^{({\rm DZ})}$) and outer ($r_{\rm out}^{({\rm DZ})}$) radii on the size of non-drifting dust grains well-mixed with gas.}
    \label{tab:dead_zone}
\end{table}

Figure~\ref{fig:x_B_ring}(a) shows that the ring appears as a region of locally decreased ionization fraction. This is due to the fact that the ionization fraction is inversely proportional to gas density $\rho=\Sigma/2H$ in the considered case. According to Figure~\ref{fig:x_B_ring}(b), the appearance of the ring in the $B_z(r)$ depends on the dust grain size. In the case of large grains, $a_{\rm d}\geq 10\,\mu$m, the magnetic field strength inside the ring increases, while it decreases in the case of smaller grains. The former effect is a reflection of the dependence $B_z\propto \Sigma$ in the ideal MHD case, as the first formula in Equation~(\ref{eq:Bz}) shows. The latter effect is a consequence of efficient magnetic ambipolar diffusion in the disk with small grains, when the dependence of the magnetic field strength on disk characteristics is more complex and is described by the second formula in Equation~(\ref{eq:Bz}). Efficient magnetic ambipolar diffusion leads to the reduction of the magnetic field strength inside the ring under the condition of small dust grains. Thus, the density ring appears as a region of decreased magnetic field strength (`magnetic gap') in the case of small grains and a region of increased magnetic field (`magnetic ring') in the case of large grains.

\section{Varying Stokes number}
\label{app:stokes}
In Section~\ref{sec:subkepler} and Figure~\ref{fig:subkepler_ring}, we studied the radial drift of the particles with a fixed value of the Stokes number $\mathrm{St}=0.01$. 
In the general case, the Stokes number is not a constant, but rather defined by particle sizes limited by the dust growth barriers. In the case of the dust fragmentation barrier~\citep[see][]{birnstiel2012}:
\begin{equation}
    \mathrm{St} = \frac{1}{3}\frac{u_{\rm frag}^2}{\alpha c_{\rm T}^2},\label{eq:stokes_frag}
\end{equation}
where $u_{\rm frag}$ is the fragmentation velocity, which is typically $u_{\rm frag}=10$~m~s$^{-1}$.

\begin{figure*}[t]
  \centering
  \includegraphics[]{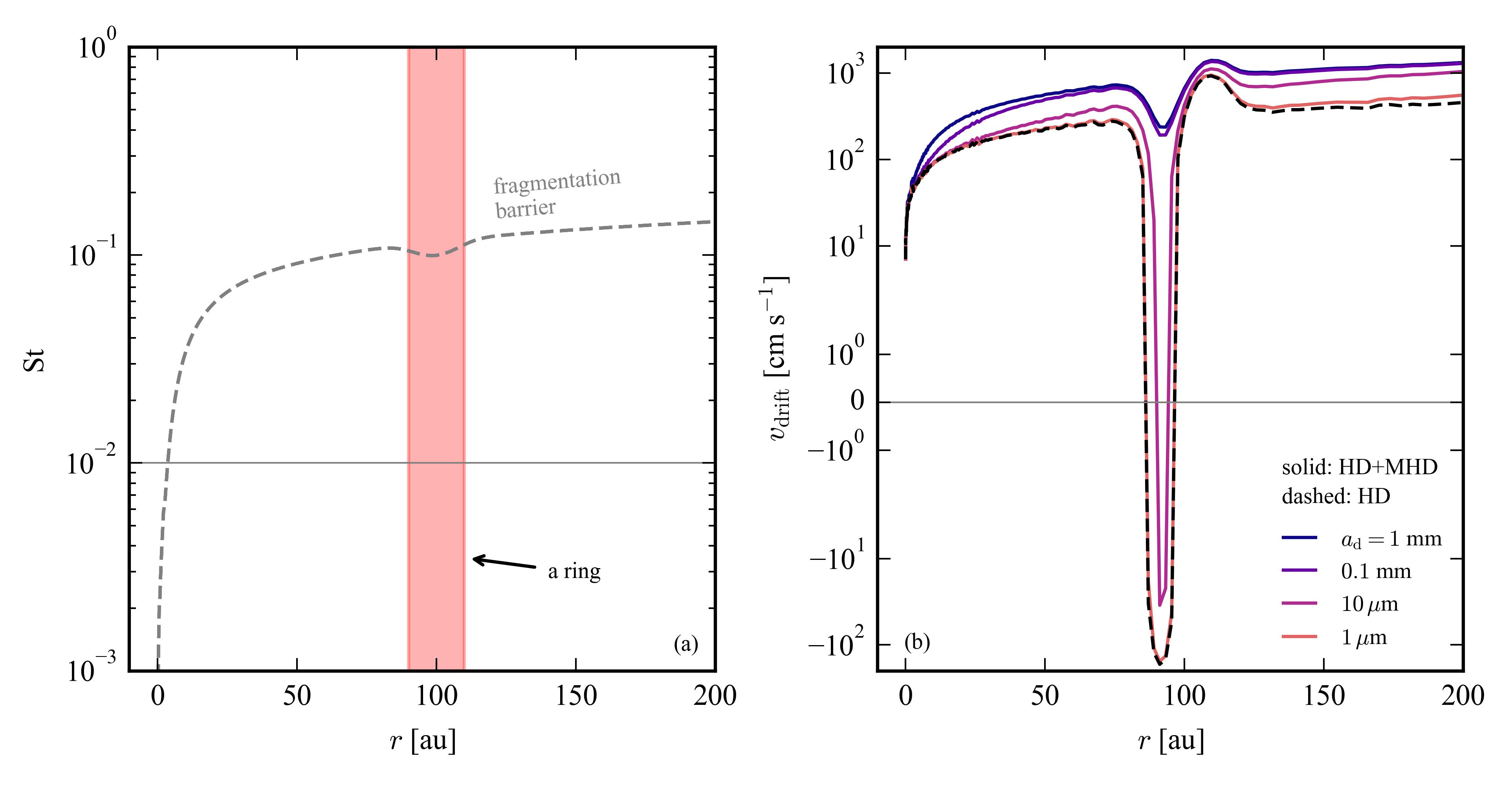}
  \caption{Panel (a): the radial profile of the Stokes determined for the drifting particles, which size is determined by the fragmentation barrier (Equation~(\ref{eq:stokes_frag}), dashed gray line). Horizontal line delineates the value $\mathrm{St}=0.01$ used as standard value in Section~\ref{sec:subkepler}. The highly turbulent disk ($\alpha=10^{-2}$) with a ring of radius $R_{\rm ring}=100$~au, half-width $10$~au, and maximum density $\Sigma_{\rm bump} = 2\times \Sigma_{\rm disk}(2) \approx 12$~g~cm$^{-2}$ is considered, as in Figure~\ref{fig:subkepler_ring}. Panel~(b): corresponding radial profiles of the stationary speed of the radial drift calculated in the cases when both HD and MHD deviations from the Keplerian rotation are taken into account (colored solid lines), and the HD deviation is considered only (black dashed line). The curves for the MHD case are shown for various sizes of non-drifting dust grains. }
\label{fig:drift_real_stokes}
\end{figure*}

In Figure~\ref{fig:drift_real_stokes}, we plot the radial profiles of the Stokes number defined as (\ref{eq:stokes_frag}), as well as the corresponding particle drift speeds. The standard disk and ring parameters are used, as in Section~\ref{sec:subkepler}.

Figure~\ref{fig:drift_real_stokes}(a) shows that the Stokes number increases with distance, ranging from $<10^{-3}$ in the inner part of the disk to $\sim 0.1$ at its outer edge. This is a direct consequence of the decreasing value of the gas turbulent speed $v_{\rm turb}\approx \sqrt{\alpha} c_{\rm T}$ due to the decreasing gas temperature. The Stokes value is small, $\mathrm{St}\ll 1$, in the inner part of the disk due to high temperature. In the bulk of the disk, $r\gtrsim 10$~au, the value of St is of $0.01-0.1$. The standard case of $\mathrm{St}=0.01$ considered in Section~\ref{sec:subkepler} presents a more conservative estimate of the fragmentation dominated regime and thus have smaller drift velocities. The change in temperature causes a little decrease in the Stokes number inside the ring. But this decrease is insignificant and does not affect the conclusions regarding the impact of the magnetic field on the drift speed.

According to Figure~\ref{fig:drift_real_stokes}(b), the behavior of the particles drift in the case of the Stokes number defined from the fragmentation limited case (\ref{eq:stokes_frag}) is qualitatively similar to the case of a fixed Stokes number. In the absence of the magnetic field (HD case), the ring acts as an efficient dust trap: the drift speed is negative inside the ring. Outside the ring, the maximum HD drift speed ranges from $10$ (inner part of the disk) to $\sim 500$~cm~s$^{-1}$ (outer part of the disk). As in the case of a fixed Stokes number, the MHD deviation from the Keplerian rotation increases the drift speed, such that the drifting particles acquire a positive drift speed inside the ring, i.e. the magnetic field of the disk opens the trap for such particles.

We performed simulations of drift speeds for particles with St number defined as (\ref{eq:stokes_frag}) and found similar results for the different disk and ring parameters. Thus, the effect of dust trap opening by magnetic tensions works universally and holds for fragmentation dominated disk as well. The value of the Stokes number determined from the fragmentation barrier exceeds the standard value $0.01$, so the effect of MHD opening of dust traps is even more prominent in this case.

\section{Dependence of the trap properties on the turbulence efficiency}
\label{app:alpha_dependence}
In Section~\ref{sect:results}, we studied the role of MHD effects on the gas rotation speed and structure of dust traps in disks with relatively high turbulence efficiency, $\alpha=10^{-2}$. Examples of such disks are DM~Tau and IM~Lup~\citep{flaherty2024, hardiman2026}. However, according to modern observational data, turbulence in many observed protoplanetary disks may be characterized by smaller values of the turbulence parameter: $\alpha\sim 10^{-3}-10^{-4}$~\citep{flaherty2020, rosotti2023}. The value $\alpha$ determines the density distribution in the disk and advection speed of the magnetic field, which in turn can change the magnetic field strength and, as a consequence, the role of MHD effects in particle drift across dust traps.

\begin{figure*}[t]
  \centering
  \includegraphics[]{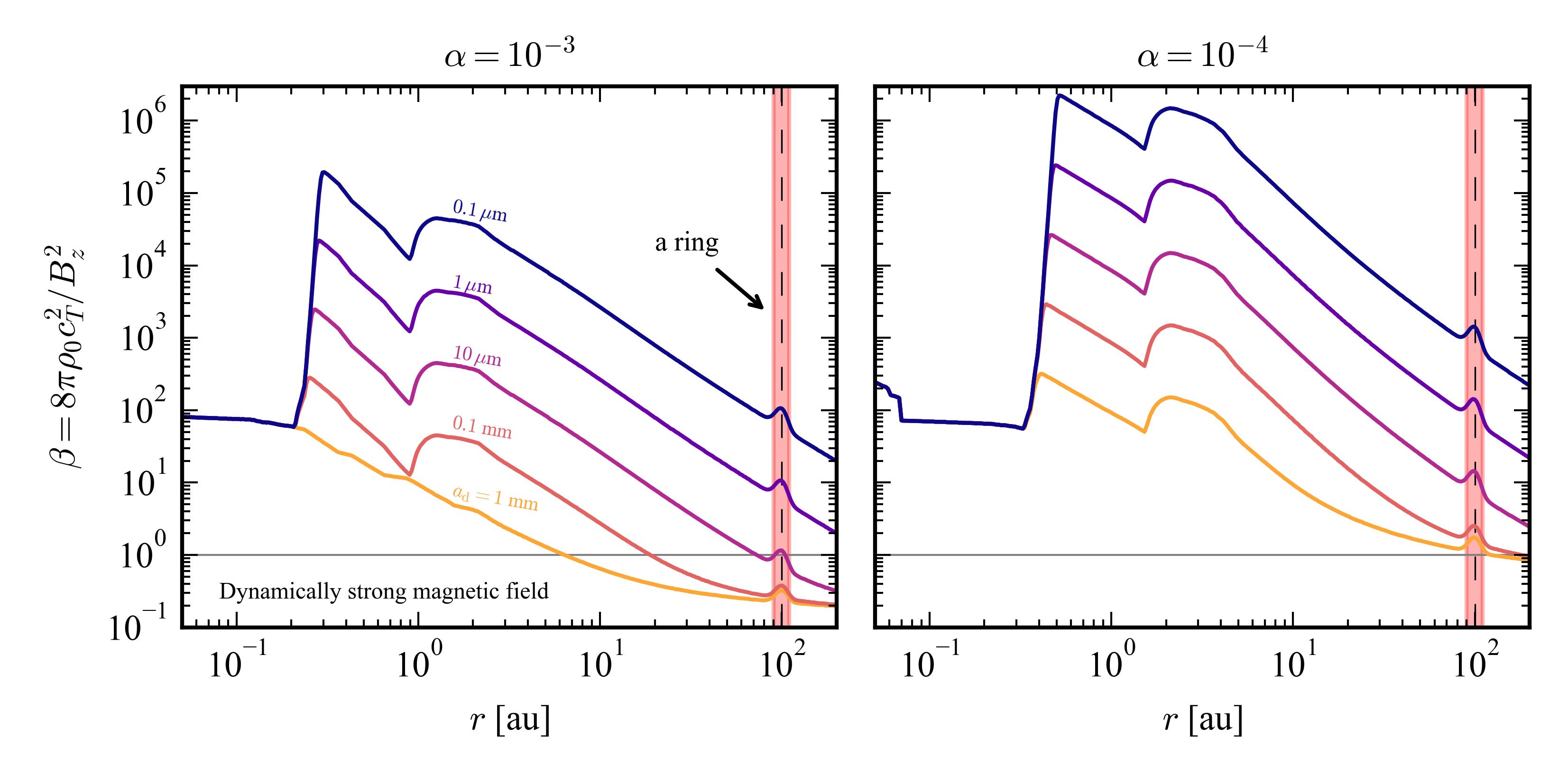}
  \caption{The radial profiles of plasma $\beta$ for various dust grain sizes (lines of different colors) in the disk with turbulence parameter $\alpha=10^{-3}$ (panel a) and $10^{-4}$ (panel b). The ring with radius $R_{\rm ring}=100$~au, half-width $\Delta R_{\rm ring}=10$~au, maximum density $\Sigma_{\rm bump}/\Sigma_{\rm disk}(R_{\rm ring})=0.8$ is considered. The horizontal line delineates value $\beta=1$. }
\label{fig:beta_low_alpha}
\end{figure*}

In this Section, we study the effect of magnetic tensions on gas rotation speed and the consequent particle drift speed in such laminar disks. To do this, we performed simulations of the disk structure for lower values of the turbulence parameter: $\alpha=10^{-3}$ (`moderate' turbulence) and $10^{-4}$ (`weak' turbulence). As an example, we consider the case of the ring with a radius $R_{\rm ring}=100$~au, half-width $\Delta R_{\rm ring}=10$~au, and maximum density $\Sigma_{\rm bump}/\Sigma_{\rm disk}(R_{\rm ring})=0.8$. The distribution of plasma beta in such disks is shown in Figure~\ref{fig:beta_low_alpha}.

Figure~\ref{fig:beta_low_alpha}(a) shows that the overall behavior of the profile $\beta(r)$ is similar to that depicted in Figure~\ref{fig:B_vs_ad}, but the values of $\beta$ are larger (i.e. the magnetic field is weaker) than in the case of $\alpha=10^{-2}$. The magnetic field is dynamically strong for $a_{\rm d}\geq 10\,\mu$m in this case, and the region of $\beta\leq 1$  lies at $r>5$~au. According to Figure~\ref{fig:beta_low_alpha}(b), the plasma $\beta$ is even larger and falls below unity near the outer edge of the disk for the case of large grains only, $a_{\rm d}=1$~mm. Thus, the decrease in $\alpha$ leads to a decrease in the magnetic field strength (in terms of plasma beta). This is a consequence of reduced accretion speed and the corresponding advection speed of the large-scale magnetic field.

\begin{figure*}[h!]
  \centering
  \includegraphics[width=0.7\textwidth]{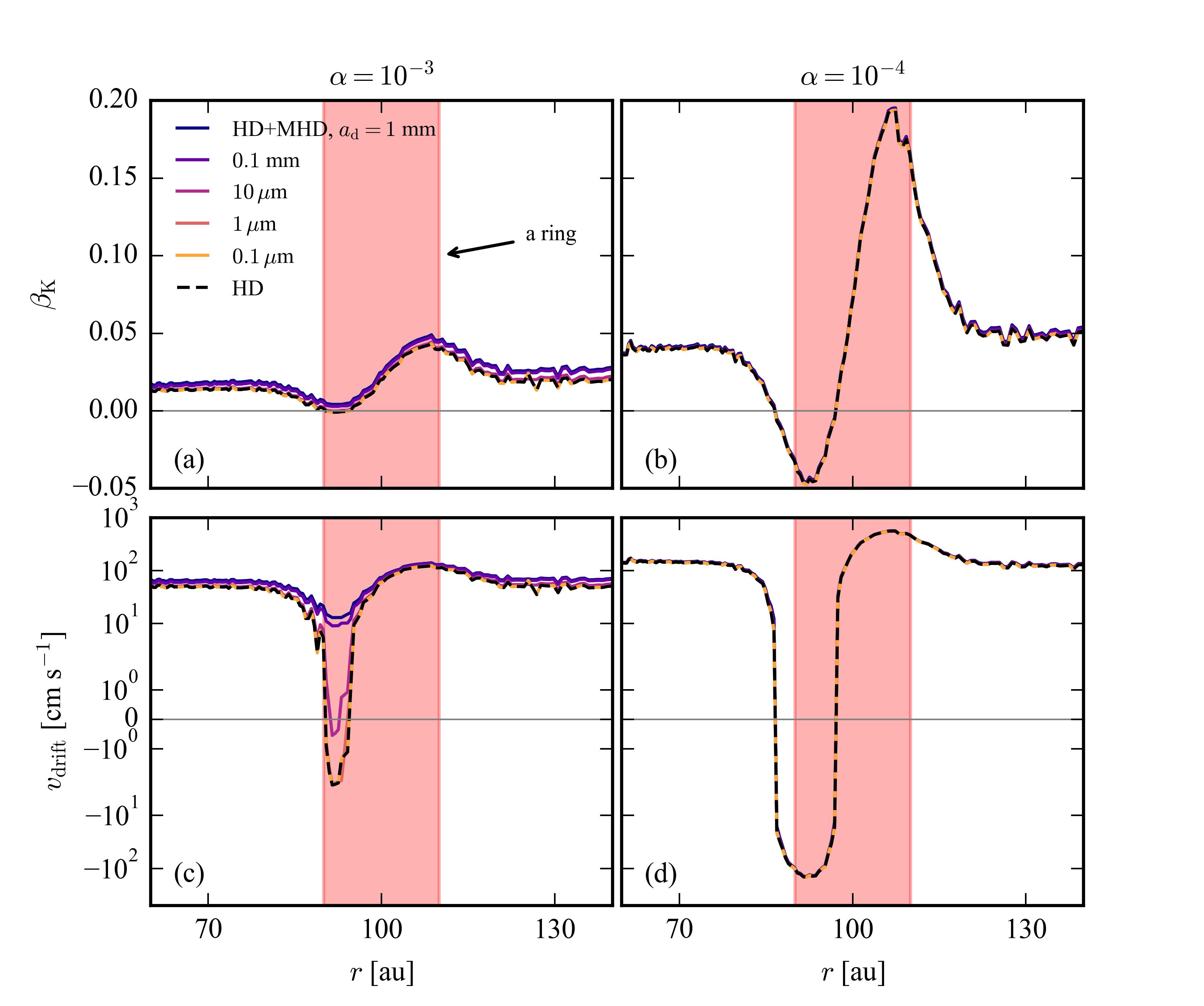}
  \caption{Dust traps properties in the disks with moderate turbulence ($\alpha=10^{-3}$, left panels) and weak turbulence ($\alpha=10^{-4}$, right pnales). The ring in the disk has a radius $R_{\rm ring}=100$~au, half-width $10$~au and a maximum density $\Sigma_{\rm bump}/\Sigma_{\rm disk} = 0.8$. Panels~(a) and (b): the radial profiles of the degree of the deviation of plasma rotation speed from the Keplerian one, $\beta_{\rm k}$, calculated for various sizes of non-drifting dust grains.  Dashed line: the HD case when the deviation is caused by a radial gas pressure gradient. Solid lines: the case when both HD and MHD deviations are considered. Panels~(c) and (d): corresponding radial profiles of the stationary speed of the radial drift for particles with Stokes number ${\rm St}=0.01$.}
\label{fig:drift_low_alpha}
\end{figure*}

In Figure~\ref{fig:drift_low_alpha}, we plot the corresponding radial profiles of the degree of deviation from the Keplerian speed and particle drift speed. Figures~\ref{fig:drift_low_alpha}(a) and (c) show that, in the case of moderate turbulence ($\alpha=10^{-3}$), the magnetic tensions cause a noticeable effect on the gas rotation speed and the drift speed of particles in the case $a_{\rm d}\geq 0.1$~mm. This effect is large enough to open the dust trap~--- the MHD deviation from the Keplerian rotation accelerates the drift speed to a value sufficient to overcome the dust trapping effect caused by a local gas pressure gradient. According to Figures~\ref{fig:drift_low_alpha}(b) and (d), the magnetic field effect on the gas rotation speed and particles drift speed in the disk with weak turbulence ($\alpha=10^{-4}$) is virtually absent. In this case, the magnetic field is dynamically weak (see Figure~\ref{fig:beta_low_alpha}(b)) and the deviation from the Keplerian rotation, as well as the particle drift speed, are determined by gas pressure gradient only.

\section{The case of a sharper ring}
\label{app:sharper_ring}
To study the conditions of the MHD trap opening, we performed simulations of the disk structure for various values of the turbulence parameter $\alpha$, the radius of the ring $R_{\rm ring}$, the half-width of the ring $\Delta R_{\rm ring}$ and the density contrast between the ring and the disk $\Sigma_{\rm bump}/\Sigma_{\rm disk}(R_{\rm ring})$.

\begin{figure*}
  \centering
  \includegraphics[]{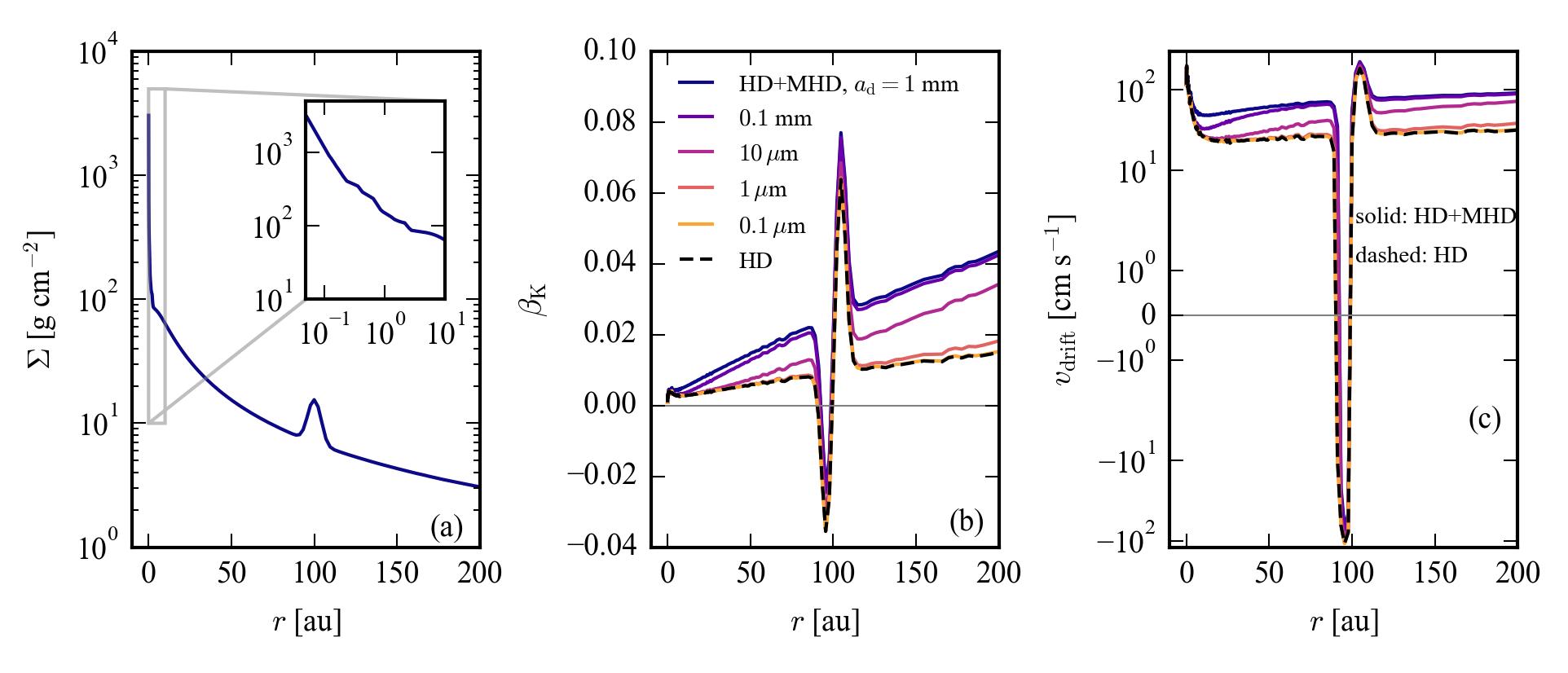}
  \caption{Same as in Figure~\ref{fig:subkepler_ring}, but for the case of a sharper ring: $R_{\rm ring}=100$~au, $\Delta R_{\rm ring}=5$~au and $\Sigma_{\rm bump}=10\,\mathrm{g}\,\mathrm{cm}^{-2}$ (see Equation~(\ref{eq:ring_density})).}
\label{fig:subkepler_ring_narrow}
\end{figure*}

As an example, Figure~\ref{fig:subkepler_ring_narrow} presents the results of the simulation of the deviations from the Keplerian rotation and the corresponding radial drift speed in the disk having a sharper ring compared to the case considered in Section~\ref{sec:subkepler} and Figure~\ref{fig:subkepler_ring}. The density contrast is the same $\Sigma_{\rm bump}/\Sigma_{\rm disk}(R_{\rm ring})=2$ as in Figure~\ref{fig:subkepler_ring}. According to observational constraints~\citep[see Introduction and][]{huang2018}, we adopted the smallest width $\Delta R_{\rm ring}=0.025\,R_{\rm ring}=2.5$~au. Figure~\ref{fig:subkepler_ring_narrow} shows that the HD deviation from the Keplerian rotation dominates over the MHD deviation for all dust grain radii considered. As a consequence, the MHD deviation does not open the dust trap caused by the radial gradient of gas pressure. However, as in the case considered in Figure~\ref{fig:subkepler_ring}, the contribution of the magnetic stresses also leads to an increase of the drift speed outside the ring.

\begin{table}[h!]
    \centering
    \begin{tabular}{|c|c|c|c|c|c|}
    \hline 
    \multirow{2}{*}{run} & \multirow{2}{*}{$\Delta R_{\rm ring}$, au} & \multirow{2}{*}{$\Sigma_{\rm bump}/\Sigma_{\rm disk}(R_{\rm ring})$} & \multicolumn{3}{|c|}{MHD opening of the trap?} \\ 
    \cline{4-6}
     &  &  & ($\alpha=10^{-2}$) & ($\alpha=10^{-3}$) & ($\alpha=10^{-4}$) \\ 
    \hline \hline
    (1) & (2) & (3) & (4) & (5) & (6) \\
    \hline
    1 & \multirow{10}{*}{$2.5$} & $10$ & no & no & no \\ 
    2 &  & $2$ & no & no & no \\ 
    3 &  & $1$ & no & no & no \\ 
    4 &  & $0.8$ & no & no & no \\ 
    5 &  & $0.6$ & no & no & no \\ 
    6 &  & $0.5$ & no & no & no \\ 
    7 &  & $0.4$ & no & no & no \\ 
    8 &  & $0.3$ & no & no & no \\ 
    9 &  & $0.2$ & yes ($a_{\rm d}\geq 10\,\mu$m) & no & no \\ 
    10 &  & $0.1$ & - & - & - \\ 
    \hline 
    11 & \multirow{6}{*}{$10$} & $10$ & no & no & no \\ 
    12 &  & $2$ & yes ($a_{\rm d}\geq 10\,\mu$m) & no & no \\ 
    13 & & $1$ & yes ($a_{\rm d}\geq 0.1$~mm) & no & no \\ 
    14 & & $0.8$ & yes ($a_{\rm d}\geq 10\,\mu$m) & no & no \\ 
    15 & & $0.6$ & yes ($a_{\rm d}\geq 10\,\mu$m) & yes ($a_{\rm d}\geq 0.1$~mm) & no \\ 
    16 & & $0.5$ & - & - & - \\ 
    \hline 
    \end{tabular} 
    \caption{Conditions of the MHD dust trap opening in disks with various ring parameters and turbulence efficiency. }
    \label{tab:dependence}
\end{table}

\section{Parameter study}
\label{app:ring_dependence}
Based on the set of performed simulations, Table~\ref{tab:dependence} presents the results indicating under which parameters the MHD deceleration of the Keplerian rotation opens the dust trap. Column $1$ gives the number of the run, columns $2$ and $3$ provide the half-width and density contrast of the ring, and columns $4-6$ list the flags indicating whether the MHD deceleration of gas rotation speed can open the trap depending on the turbulence parameter $\alpha$. Label `no' means that the trap cannot be opened by the magnetic field, `yes' corresponds to a positive condition (with the corresponding range of dust grain radii in brackets), and label `-' stands for the case of no dust trap (gas pressure gradient is too small to alter the direction of the drift).

The table shows that the dust traps inside narrow rings (minimum possible $\Delta R_{\rm ring}=2.5$~au according to observational data) can be opened only in the highly turbulent disks ($\alpha=10^{-2}$) with rings having a small density contrast of $20$~\% (column $4$, row $9$ in Table~\ref{tab:dependence}) and dust grains of radius $\geq 10\,\mu$m. The rings with higher contrast (rows $1-8$) are controlled by the gas pressure gradient. The rings with smaller contrast (row $10$) do not have traps at all due to a very small local gas pressure gradient. The disks with moderate ($\alpha=10^{-3}$) and weak ($\alpha=10^{-4}$) turbulence do not have a sufficiently strong magnetic field capable of opening the trap.

The dust traps inside wide rings (maximum possible $\Delta R_{\rm ring}=10$~au according to observational data) in highly turbulent disks ($\alpha=10^{-2}$) can be opened in the case of a  density contrast smaller than $200$~\% (column $4$, rows $12-15$ in Table~\ref{tab:dependence}). The rings with a density contrast smaller than $50$~\% do not have traps. The wide rings in the disks with moderate ($\alpha=10^{-3}$) turbulence can be opened  in the case of a density contrast of $60$~\% only (column $5$, row $15$). The magnetic field in the disks with weak turbulence ($\alpha=10^{-4}$, column $6$) is too small to open the trap.

\bibliography{lit.bib}{}
\bibliographystyle{aasjournalv7}

\end{document}